\documentclass{aa}  

\usepackage{graphicx}
\usepackage{txfonts}
\usepackage{hyperref}
\begin{document}

   \title{Stacked Bayesian general Lomb-Scargle periodogram: Identifying stellar activity signals}

   \author{A. Mortier\inst{1}
          \and
          A. Collier Cameron\inst{1}
          }

   \institute{$^1$ Centre for Exoplanet Science, SUPA, School of Physics and Astronomy, University of St Andrews, St Andrews KY16 9SS, UK\\
        \email{am352@st-andrews.ac.uk}
             }

   \date{Received ; accepted }

 
  \abstract
   {Distinguishing between a signal induced by either stellar activity or a planet is currently the main challenge in radial velocity searches for low-mass exoplanets. Even when the presence of a transiting planet and hence its period are known, stellar activity can be the main barrier to measuring the correct amplitude of the radial velocity signal. Several tools are being used to help understand which signals come from stellar activity in the data.}
   {We aim to present a new tool that can be used for the purpose of identifying periodicities caused by stellar activity, and show how it can be used to track the signal-to-noise ratio (SNR) of the detection over time. The tool is based on the principle that stellar activity signals are variable and incoherent.}
   {We calculate the Bayesian general Lomb-Scargle periodogram for subsets of data and by adding one extra data point we track what happens to the presence and significance of periodicities in the data. Publicly available datasets from HARPS and HARPS-N were used for this purpose. Additionally, we analysed a synthetic dataset that we created  with SOAP2.0 to simulate pure stellar activity and a mixture of stellar activity and a planet.}
   {We find that this tool can easily be used to identify unstable and incoherent signals, such as those introduced by stellar activity. The SNR  of the detection grows approximately as the square root of the number of data points, in the case of a stable signal. This can then be used to make decisions on whether it is useful to keep observing a specific object. The tool is relatively fast and easy to use, and thus lends itself perfectly to a quick analysis of the data.}
   {}


   \maketitle
   \titlerunning{Stacked Bayesian general Lomb-Scargle periodogram}

\section{Introduction}

The radial velocity (RV) technique is the second most successful method for discovering exoplanets, accounting for about $20\%$ of the discoveries\footnote{For accurate numbers, see \url{www.exoplanet.eu}}. Furthermore, RV measurements are being used to follow up planets discovered by the transit method to determine the bulk densities of transiting planets and to look for more non-transiting planets in the system \citep[see for example][]{Buch16}.

Periodic changes in the RV of a star can provide information on the planets orbiting that star, such as their period and mass. However, there are other effects that may introduce periodic variations in the measured RV of a star, of which the most common ones are instrumental effects, time sampling effects, and the intrinsic variability of the star. 
Although we can improve instrument stability and optimise observing schedules, we cannot control the behaviour of the stars we observe. Stellar variability is currently the main barrier to detecting true Earth-like planets. Determining the significance and nature of the signals seen in the data is key to finding and characterising exoplanets with the RV method.

Several techniques exist to try and distinguish whether a periodic signal comes from stellar activity or from planets. These techniques are often based on the fact that periodic RV variability induced by a planet is stable and coherent over time and across all wavelengths whereas stellar-activity-induced variations are not. These variations could have a single underlying period, such as the stellar rotation period, but its phase, amplitude, and shape may evolve with time owing to the transient nature of stellar activity \citep[for an overview of different stellar activity signals see e.g. ][]{Dum12}.

Spectroscopic observations of the same star can be made in multiple wavelengths. If the RV amplitude is not the same in all wavelengths, the variation is not caused by a planet but rather by a stellar spot \citep[e.g. \object{TW Hya} -][]{Hue08}. Because a spot is cooler than the photosphere, and assuming black body radiation, a spot peaks in longer wavelengths than the photosphere. This makes the flux contrast between spot and photosphere smaller for longer wavelengths. A spot thus blocks less light in longer wavelengths, decreasing the RV variability.

Spots and faculae have a different brightness. Photometry of the star made simultaneously with spectroscopic observations can then shed light on the amount of stellar variability and its periodicity during the observations. This has already been done in some cases, for example for \object{HD166435} \citep{Que01}. Recently, \citet{Aig12} and \citet{Hay14} went a step further and used photometric information in the analysis of the RVs of \object{HD189733} and \object{CoRoT-7}.

Another standard practice is to use activity indicators by investigating the presence of any correlations of the indicators with the RV . If the RV variations are caused by the presence of a planet, they should not correlate with activity-sensitive measurements. If the RVs are determined using a cross correlation function (CCF), the properties of the CCF can be used as activity tracers, such as the bisector span or full width at half maximum \citep[e.g.][]{Que01, San14, Raj15}. Specific lines in the spectrum could also be used to trace activity, such as the \ion{Ca}{II} H\&K emission or the H$\alpha$ line \citep[e.g.][]{San14, Rob15b, Raj15}.

In this paper we explore another tool to distinguish between coherent and incoherent periodic signals. We describe the principle in Sect. \ref{SecPrin} and test it on some existing data in Sect. \ref{SecTest}. In Sect.4, we show how white noise, data sampling, and alias signals affect the results. Section \ref{SecSNR} details how this tool can be used to track the significance of the detection and we conclude in Sect. \ref{SecCon}.

\section{Stacking periodograms}\label{SecPrin}

Because the RV variations induced by a planet are periodic, it is standard practice to first run a periodicity analysis on the data when looking for planetary signals. Several tools are available to perform this sort of an analysis, such as a Fourier transform, the Lomb-Scargle periodogram \citep[LS -][]{Lomb76,Sca82}, or the general Lomb-Scargle periodogram \citep[GLS -][]{Zech09}. In this work, we make use of the Bayesian GLS periodogram \citep[BGLS -][]{ME15}.

In theory, the power or probability of a coherent long-lived signal in the data should increase by adding more observations \citep[e.g. ][]{How11,Hat13,Sua16}. On the other hand, if there is a quasi-periodic signal with varying amplitude and phase (an incoherent signal) present in the data, as is the case for activity-induced signals, the power or probability can either increase or decrease at different epochs as more data are added. We would like to caution the reader that whilst a signal whose significance fluctuates is a sign of activity, an increasing signal is not necessarily a sign of a planet. 
Additional tests should always be performed to establish the planetary nature of a signal.

We designed a tool to make the BGLS periodogram from the first $n$ points of the data. We then add the next data point, recalculate the BGLS periodogram, and repeat multiple times. We normalise all BGLS periodograms with their respective minimum values. In this way we can compare how a signal gets stronger (or weaker) over time.

We can then plot the number of data points versus the period (or frequency), and colour code the points with the probability of the BGLS periodogram. The x-axis is then the same as we see for any periodicity tool. The y-axis can be seen as a time axis and a signal is tracked vertically in the diagram. It is worth noting that calculating periodograms for subsets of data has been used before by other authors to check the presence and significance of signal periodicities over time \citep[e.g. ][]{Bru14, Ram16}.

\section{Testing the tool}\label{SecTest}

To test this tool, we looked for stars where it has been shown that there is significant RV variation due to the stellar activity. For this reason we chose to use RV data from the Sun and CoRoT-7. Furthermore, we used data for \object{Gl 581} where planet b is very well detected. As a final proof of concept we used a purely synthetic dataset.

\subsection{Gl 581}\label{secGl581}

\begin{figure}
\centering
\includegraphics[width=\hsize]{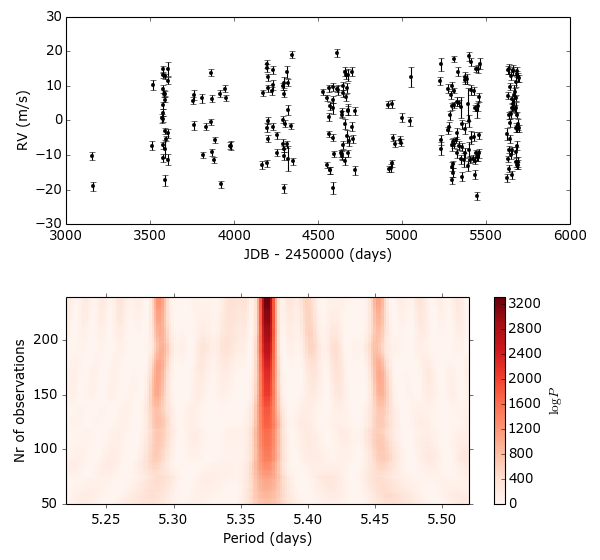}
\caption{Top: RVs from Gl581, taken by HARPS. Bottom: Stacked BGLS periodogram for the data in the top plot, zoomed around 5.4 days. Amount of observations is plotted against period, with the colour scale indicating the logarithm of the probability, where darker is more likely.}
\label{FigGl581}
\end{figure}

Gliese 581 is a close M dwarf that has been studied extensively over the years. In 2005, it was announced that this star was orbited by a Neptune-mass planet at an orbit of about 5.35 days \citep{Bon05b}. This result was based on data from HARPS at the 3.6m telescope in La Silla, Chile \citep{Mayor03}. The same survey team subsequently announced the discovery of three additional planets in the system \citep{Udry07b,Mayor09}. In 2010, \citet{Vogt10} announced the presence of six planets in the system using the HARPS data and additional data from HIRES at Keck I in Hawaii \citep{Vogt94}. This announcement was particularly interesting because it placed Gl581g in the habitable zone of the planet.

Several of the announced signals are now suspected to be related to stellar activity rather than planets \citep[e.g.][]{Forv11,Rob14b} and we  discuss this in Sect. \ref{SecSNR}. The signal from Gl581b, however, is a very clear and strongly-detected signal. This makes it a good test case for our tool to show how a signal from a planet should behave as more observations are added. In this paper we used the HARPS data that was published in \citet{Forv11} which has 240 data points, spanning about seven years of observations.

Figure \ref{FigGl581} shows the data and the stacked BGLS periodogram. The signal at 5.37\,days is very well-detected. With all observations included, this signal has a probability of more than $10^{3000}$ (with the minimum probability set to $1$). Furthermore, the signal steadily becomes more significant by adding observations. This is exactly what we expect for a signal that arises from the presence of a planet.

\subsection{The Sun}\label{SecTestSUN}

The Sun is an ideal test case for studying stellar activity. Not only is it the only star we can resolve, but the orbits of the Solar System planets are also well defined. As such, we can predict the orbital reflex motion of the Sun, which is induced by the planets, at any given time . By subtracting this, we are left with RV data of a G-type star of which we are certain that all the variations are caused by the star itself.

\begin{figure}
\centering
\includegraphics[width=\hsize]{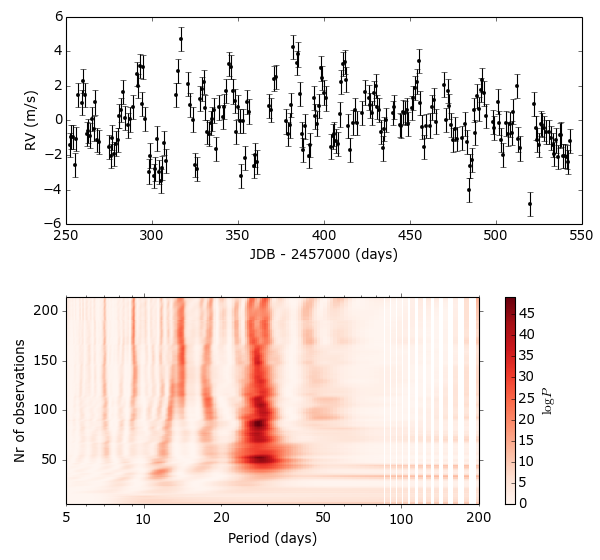}
\caption{Top: RVs from the Sun, taken by HARPS-N. The data was binned per day. Bottom: Stacked BGLS periodogram for the data in the top plot, colours as in Fig. \ref{FigGl581}.}
\label{FigSUN}
\end{figure}

We used the data from \citet{Phi17}. These RVs were taken via a solar telescope connected with HARPS-N at the TNG in La Palma \citep{Cos12,Dum15,Phi16}. Data were taken with exposures of five minutes during daylight. Details on the data reduction can be found in \citet{Phi17}. Because we are interested in signals around the solar rotation period (approximately 27 days), we binned the data per day. This left us with 215 data points.

Figure \ref{FigSUN} shows the RV data and the stacked BGLS. A significant signal can be seen growing stronger and weaker over time, corresponding to an unstable quasi-periodic signal such as stellar activity variations. The main signal is seen around 27 days, corresponding to the solar rotation period. When this signal becomes weaker, another signal at around 13.5\,days becomes stronger. The latter signal is the harmonic of the solar rotation period, $P_{rot}/2$.

\subsection{CoRoT-7}

CoRoT-7 is another star that has been debated in the literature. This G dwarf is orbited by a transiting planet at 0.85\,days and a non-transiting planet at 3.7\,days and exhibits significant activity signals, as is obvious from the light curves \citep{Que09}. \citet{Hat10} claimed that there was a third planet, CoRoT-7d, with a period of 9.02 days. \citet{Hay14} showed that the 9-day signal arose most likely from stellar activity. They used spectroscopy data from HARPS and a Gaussian-process regression model of the stellar activity based on simultaneous photometry from the CoRoT spacecraft \citep{Barr14} in their analysis.

\begin{figure}
\centering
\includegraphics[width=\hsize]{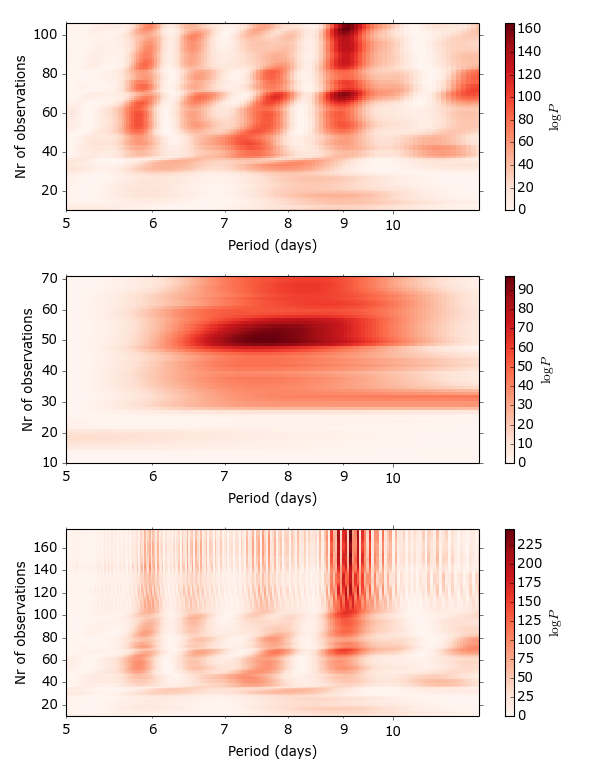}
\caption{Stacked BGLS periodograms for CoRoT-7, colours as in Fig. \ref{FigGl581}. Top: using RVs from HARPS by \citet{Que09}, middle: using RVs from HARPS by \citet{Hay14}, bottom: using the combined dataset.}
\label{FigC7}
\end{figure}

We use the available HARPS RVs from \citet{Que09} and \citet{Hay14}. Figure \ref{FigC7} shows that the 9-\,day signal was already unstable in the original dataset from \citet{Que09}, seen in the top plot. The second dataset, from \citet{Hay14}, also shows the unstable nature of this signal around 9 days (middle plot). It is interesting to point out that the 9-day signal  is a lot broader for the second dataset than for the first one. This is due to the limited timespan of the data (25 days versus 109 days). 

The combined dataset again shows that the 9-day signal is not stable over time, making its origin more likely to be stellar activity than a third planet. The fringes that appear in the stacked BGLS periodogram of the combined dataset arise from uncertainty in the cycle count across the three-year gap separating the 2009 and 2012 RV campaigns.

\subsection{Synthetic data}\label{secsyn}

\begin{figure*}
\centering
\includegraphics[width=\hsize]{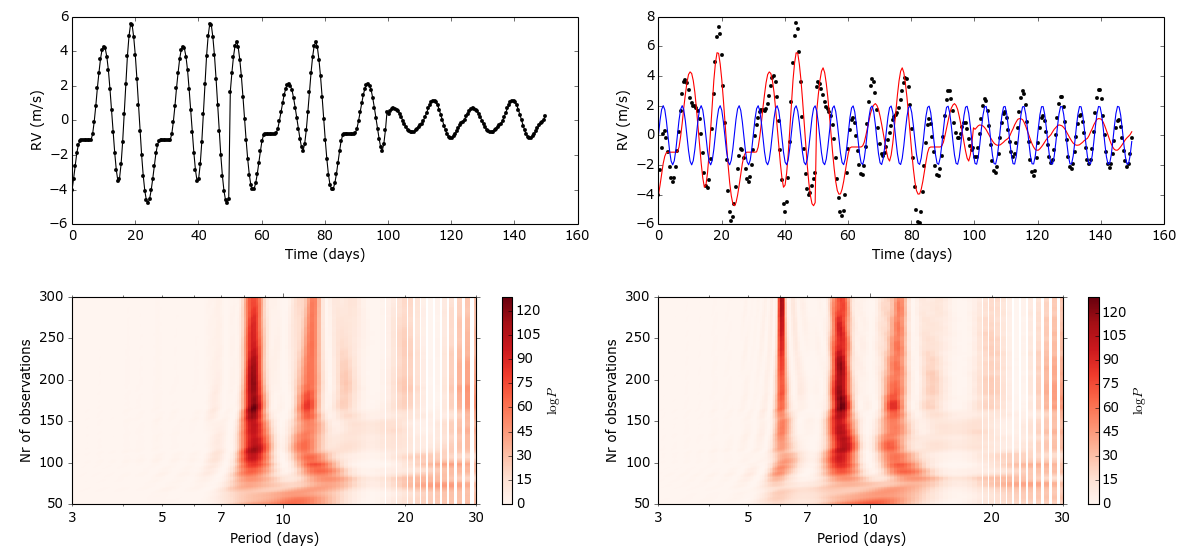}
\caption{Top left: Synthetic RVs generated by SOAP2.0. Top right: Black points are the combination of a stellar-activity signal generated by SOAP2.0 (red curve) and a signal of a planet on a six-day circular orbit (blue curve). Bottom: Stacked BGLS periodogram for the data in the top plots, colours as in Fig. \ref{FigGl581}.}
\label{FigSynTest}
\end{figure*}

As a proof of concept, we also created  and analysed a synthetic dataset. To generate RVs related to stellar activity, we used SOAP2.0 \citep[Spot Oscillation and Planet][]{Dum14b}. This tool estimates photometric and RV variations induced by stellar spots and plages. It is an upgrade from its older version SOAP \citep{Boi12} and includes convective blueshift, limb brightening of plages, and more realistic stellar physics.

SOAP2.0 does not allow for spot evolution, therefore we created three separate sets of spot configurations that we then concatenated to create a dataset of a star with evolving active regions. The star we simulated has properties like the Sun, with a radius of one Solar radius, an effective temperature of 5778\,K, a spot temperature of 5115\,K, quadratic-law limb-darkening coefficients $c_1=0.29$ and $c_2=0.34$, and a rotation period of $25.05$\,days. We observed the star equator-on (inclination $i= 90^{\circ}$) with a simulated spectrograph of resolution comparable to HARPS(-N), 115000.

The three spot configurations each consisted of two spots. The first set had spot radii of $10\%$ and 14$\%$ of the stellar radius, at latitudes of $30^{\circ}$ and $45^{\circ}$ and longitudes of $180^{\circ}$ and $60^{\circ}$. The consequent sets had spot radii of twice $10\%$ and twice $5\%$, latitudes of $50^{\circ}$, $30^{\circ}$, $45^{\circ}$, and $30^{\circ}$ and longitudes of $60^{\circ}$, $300^{\circ}$, $300^{\circ}$, and $120^{\circ}$. We used each configuration for two full rotation periods and sampled 50 points per rotation period. In total we ended up with a dataset of 300 points with a timespan of 150.3 days. We used no additional noise model and assumed uniform errorbars in our analysis.

The left panel of Fig. \ref{FigSynTest} shows the synthetic data and its stacked BGLS periodogram. Two periodicities show up at about $12.5$\,days and $8.35$\,days, respectively $P_{rot}/2$ and $P_{rot}/3$. Both signals grow stronger and weaker over time as expected.

We added a sinusoid to the data, simulating a planet on a circular six-day orbit with a semi-amplitude of $2$\,m/s. This combined dataset is plotted in the top right panel of Fig. \ref{FigSynTest}. The stacked BGLS periodograms still pick up the varying periodicities at $12.5$ and $8.35$ days. At the same time, there is a periodicity that steadily becomes more significant over time at 6 days. This shows that the stable and coherent signal is still picked up even if the stellar activity signal has similar amplitudes as the planetary signal.

The most difficult form of stellar activity signals to analyse are the ones that arise from long-lived active regions. Because these signals stay coherent over long timescales, it is easier to confuse them for real planetary signals. To see the effect of long-lived active regions, we simulated a spot on the same star as before, with a radius of 
$14\%$ the stellar radius at a latitude of $45^{\circ}$. We ran SOAP2.0 for five rotation periods, changing neither the amplitude nor phase of the active region. All other input parameters stayed the same as the first simulation.

\begin{figure}
\centering
\includegraphics[width=\hsize]{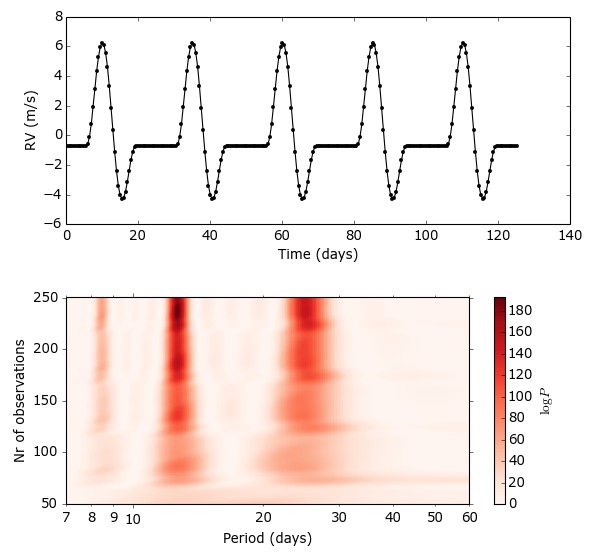}
\caption{Top: Synthetic RVs generated by SOAP2.0. Bottom: Stacked BGLS periodogram for the data in the top plot, colours as in Fig. \ref{FigGl581}.}
\label{FigSynTest2}
\end{figure}

Figure \ref{FigSynTest2} shows the resulting RV data and its corresponding stacked BGLS periodogram. The three periodicities related to the rotational period clearly show up. Although it is not as clear-cut as in other examples, we can still see the main periodicity (at $12.5$\,days) varying in significance over time, distinguishing itself from a truly coherent sinusoid.

\section{Exploring the possibilities}\label{SecPoss}

In this section we demonstrate that periodicity analysis tools can identify periods that are solely due to the sampling of a dataset \citep[e.g. ][]{Raj16}. Significant peaks can arise at periods that are also present in the window function. Signals also have alias signals due to the observation schedule. 

\subsection{Effect of data sampling}

\begin{figure}
\centering
\includegraphics[width=\hsize]{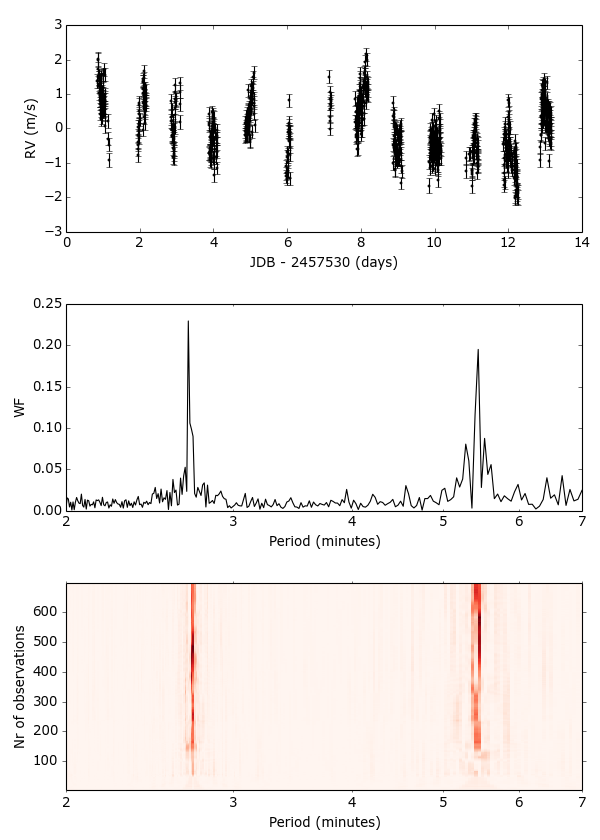}
\caption{Top: RVs from the Sun, taken by HARPS-N. Middle: Window function of the data, zoomed around 4 minutes. Bottom: Stacked BGLS periodogram for the data in the top plot, zoomed around 4 minutes, colours as in Fig. \ref{FigGl581}.}
\label{FigWF}
\end{figure}

Ground-based observations suffer from having gaps in the data due to reasons such as the telescope not being available and the weather. Even with continuous access to a telescope and an instrument, and perfect weather, there is still be a day-night cycle that automatically creates gaps in the data. Furthermore, observations have finite exposure times and a data point is thus always an average over a small time frame. These gaps and finite exposure times introduce periodicities in the data that are only due to these effects.

It is thus important that the window function of the data is also explored. Significant periodicities in the window function will most likely also show up in the periodicity analysis of the data although they are not caused by any stellar or planetary effect. We can show this fairly clearly by using the solar data. Observations were taken with an exposure time of 5 minutes. Including the time for read out and starting the new observation, the data points were eventually spaced by $5.4$ minutes.

We used thirteen consecutive days of data from \citet{Phi17}, shown in the top plot of Fig. \ref{FigWF}. For calculating the window function, we used Equation 8 in \citet{Rob87}. When zoomed in around 5 minutes, the window function shows two peaks at $5.4$ and $2.7$ minutes, the latter being half the period of the first. These peaks are solely due to the spacing of the data and the finite exposure time.

When we calculate the stacked BGLS periodogram, as shown in the bottom plot of Fig. \ref{FigWF}, we see that these periods also arise. They are not stable, most likely because there are other time spacings in the data as well.

\subsection{Effect of alias signals}

\begin{figure}
\centering
\includegraphics[width=\hsize]{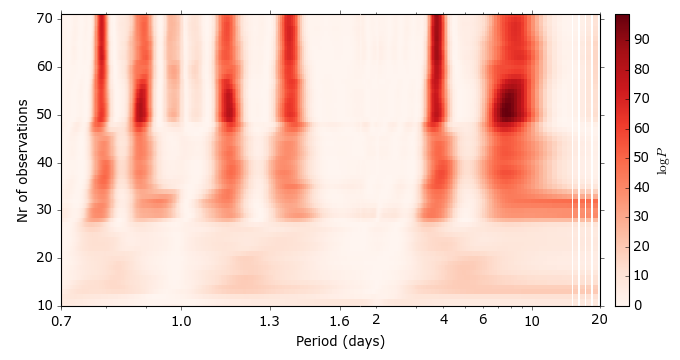}
\caption{Stacked BGLS periodogram for CoRoT-7, colours as in Fig. \ref{FigGl581}. The same data is used as for the middle plot of Fig. \ref{FigC7}.}
\label{FigAlias}
\end{figure}

Alias signals arise very often in RV datasets due to the way our data is sampled and the typical properties of ground-based observations. In most cases, telescopes are shared between several groups and there is no continuous access to take the data one needs. Furthermore, there are the natural cycles we deal with such as the day-night cycle, the seasonal cycle of visible stars in the sky, or the lunar cycle. Recognising the signals that are aliased is thus an important step in the data analysis \citep[e.g. ][]{Daw10}.

Because aliased signals are linked to the true signal and often come from the same natural constraints of our observations, they behave similarly in terms of significance. In the stacked BGLS periodogram, we can then see if there are periodicities that behave similarly. We used the CoRoT-7 data from \citet{Hay14} for this purpose.

We show the results in Fig. \ref{FigAlias}. There is an increasing signal around 3.7 days (coming from CoRoT-7c) and the activity signal around 9 days. These signals are aliased due to the 1-day sampling frequency of ground-based observations. The alias periods are then calculated with $P_{alias} = 1/|1/P_{real} \pm 1|$ \citep{Dum12}. Figure \ref{FigAlias} shows that both the signal from 3.7 days and the 9-day signal appear around 1 day as well. They are slightly less significant than the true signal\footnote{Note that is not always the case for alias signals.}, but more importantly they exhibit a similar behaviour overall.

\section{Tracking the significance of detection}\label{SecSNR}

Stacking periodograms can be used for more than tracking the significance of a periodicity. It can also track the significance (or SNR) of the semi-amplitude for a sinusoid at a particular period . 

\subsection{Principle}

\begin{figure*}
\centering
\includegraphics[width=\hsize]{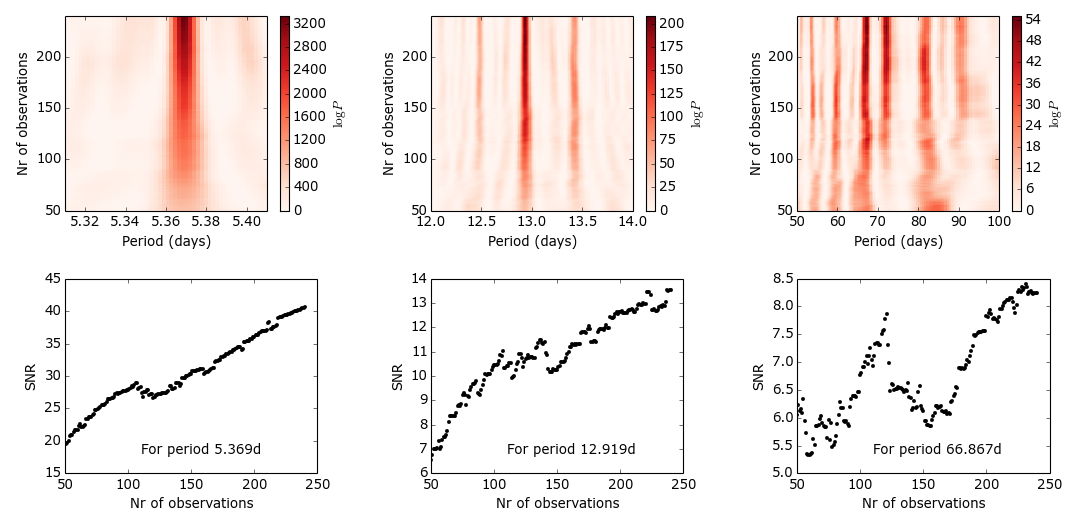}
\caption{Top: Stacked BGLS periodograms of Gl581, colours as in Fig. \ref{FigGl581}. The same data as Fig. \ref{FigGl581} was used. Bottom: $SNR_K$ versus amount of observations for the most significant periodicity. Left to right: Full dataset, subtracted best fit at 5.37 days, subtracted best fit at 12.9 days.}
\label{FigSNR}
\end{figure*}

A periodogram essentially finds the best solution for fitting a sine curve at every period, by using a maximum likelihood estimation or least squares minimisation . Alongside the probability (or power) of each period, it thus also provides the best-fitted semi-amplitude $K$ and phase $\phi$ for each period. The significance of a detection at a particular period can be expressed as $K/\sigma_K = SNR_K$ with $\sigma_K$ being the error bar on the fitted parameter. This error bar can be estimated using optimal scaling.

We assume the data is fitted by a sine curve\footnote{Note that we set the offset to zero here for simplicity. It is straightforward to include it.}: 

\begin{equation}
d(t_i) = K\sin\left(\frac{2\pi}{P} t_i + \phi\right) = K\sin(\varphi_i),
\end{equation}

\noindent where $K$ is the semi-amplitude, $P$ the orbital period, and $\phi$ the phase. Each data point $d_i$ taken at time $t_i$ has an associated error bar $\sigma_i$. Through minimisation of chi-squared, weighted by the inverse variance, we can then find that

\begin{equation}
K = \frac{\sum_i w_i d_i \sin\varphi_i}{\sum_i w_i \sin^2\varphi_i}
\end{equation}

and 

\begin{equation}
\sigma_K = \frac{1}{\sum_i w_i \sin^2\varphi_i},
\end{equation}

where $w_i = 1/\sigma_i^2$ are the weights assigned to the data points. For a particular period, we can thus calculate this detection significance $SNR_K$ for each set of observations and track how this significance evolves. At the same time, the value for $K$ can be followed. This value should stabilise by adding more observations, if the periodic signal is induced by a planet.

Most radial velocity time series contain more variation than just one sine curve, either through additional planetary signals or through stellar-induced signals. To account for this extra variation, we include a white noise term $s$ in our modelling so that $w_i = 1/(\sigma_i^2+s^2)$. For each set of observations we determine the best white noise $s$ by calculating the BGLS likelihood for a range of different values of $s$ and choosing the $s$ of the maximum likelihood.

\subsection{Test on Gl 581}

To test the tracking of the semi-amplitude, we used the HARPS data of Gl 581, as described in Sect. \ref{secGl581}. Figure \ref{FigSNR} shows the stacked BGLS periodogram on the top, and the semi-amplitude $K$ and its detection significance $SNR_K$ on the bottom. Additionally, we pre-whitened the data to examine the remaining signals in the data. Using all observations, the best-fitted model for one planet on a circular orbit is selected, including its white noise $s$. This model is then subtracted from the data.

Planets b and c at periods $5.37$\,days and $12.9$\,days show strong and increasing signals . The SNR of the detection steadily grows by adding more observations, reaching a $40\sigma$ and $13\sigma$ detection respectively. This is in good agreement with the values found in the literature. The next significant periodicity is at about $66.8$\,days. The nature of this signal is being disputed in the literature, with some authors attributing the signal to a planet \citep[e.g.][]{Mayor09,Vogt10,Angl15} and other authors attributing it to stellar activity \citep[e.g.][]{Balu13b,Rob14b}.

Figure \ref{FigSNR} shows that although the signal strength grows by adding more observations, it is not doing so in a steady manner. The signal detection significance varies between a $6\sigma$ and $8\sigma$ detection. This shows that there is some variation in the signal over time despite it being a fairly significant detection. Ascertaining the true nature of this signal will require more testing and better modelling techniques, but it is clear from this straightforward analysis that the signal should be modelled with care.

\subsection{Test on synthetic data}

\begin{figure*}
\centering
\includegraphics[width=\hsize]{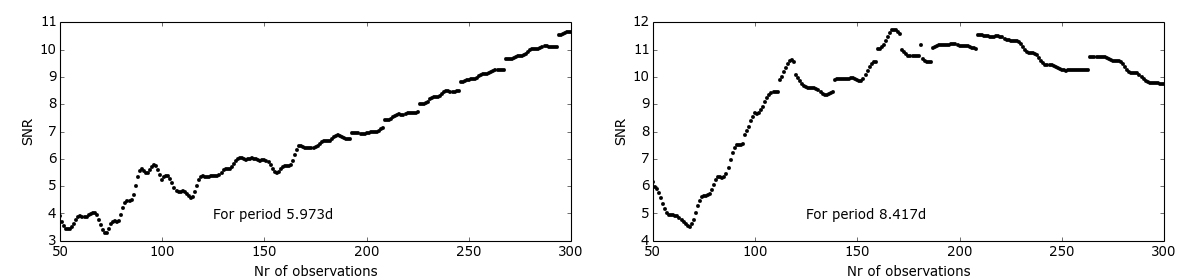}
\caption{Left: $SNR_K$ versus amount of observations for the most significant periodicity in the synthetic dataset plotted top right in Fig. \ref{FigSynTest}. Right: Same figure as the left panel, but the best-fitted signal at 5.973 days has been subtracted from the data.}
\label{FigSynSNR}
\end{figure*}

We also used the synthetic dataset from Sect. \ref{secsyn}, with the planet signal included, to showcase how the detection significance $SNR_K$ changes for a true coherent and true uncoherent signal. Results are shown in Fig. \ref{FigSynSNR}.

The best period in the complete dataset is found to be $5.973$\,days, very close to the true period of $6$\,days. The $SNR_K$ grows steadily by adding more simulated observations as expected for a coherent stable signal. The best circular fit for the complete dataset was found with a semi-amplitude of $1.87\pm0.18$\,m/s and white noise of $1.9$\,m/s. This is, within error bars, the same as the true semi-amplitude of $2$\,m/s.

After subtracting the best fit from the data, we did the analysis again. The most significant periodicity is now $8.417$\,days, relating to one third of the rotation period. The detection significance $SNR_K$ first grows steadily but then flattens off and even begins to drop when more observations are added. This is a clear sign of an incoherent signal, which is the true nature of this simulated signal.

\section{Discussion and conclusions}\label{SecCon}

In this paper we present a new tool that can be used for the purpose of identifying periodicities caused by stellar activity. The tool is based on the BGLS periodogram and the principle that planetary signals are stable and coherent whereas stellar activity signals are not. We stack the periodograms by subsequently adding observations and recalculating the periodogram. 

We used public datasets from Gl 581, CoRoT-7, and the Sun taken with HARPS and HARPS-N, and synthetic datasets using SOAP2.0. We show that the signals attributed to planets are coherent and grow steadily more significant with time. Signals that are known to come from stellar activity (the Sun, and the synthetic data), or found to be activity-related through other methods (CoRoT-7), clearly show up as being unstable over time in our analysis. On the other hand, a true coherent sinusoid injected in the synthetic data is easily picked up as becoming more significant by adding observations. Stacking periodograms reveals that aliased signals behave similarly to their true signals, making it easier to spot related periodicities.

Because the BGLS periodogram fits a sinusoid for each period, it provides a semi-amplitude as well. We estimated the error bar on the semi-amplitude through optimal scaling. We used this to calculate the significance of detection for a chosen periodicity. Tracking this detection significance $SNR_K$ could, for example, be useful for deciding if it is worth spending more precious telescope time on an object.

Determining the nature and origin of periodic signals in radial velocity datasets has become more important in our hunt for a true Earth-like exoplanet. Many different techniques exist for these types of analyses, including Gaussian process regression and apodised Kepler periodograms \citep[e.g.][ and references therein]{Hay14,Raj15,Greg16,Dum17}. This tool complements these existing techniques and can be used as an additional test in the data analysis. Stacking periodograms is a fast and intuitive way to visualise how the data and its underlying periodicities are behaving over time.

Given that we only assume one sinusoid in our underlying model, there could be cases where this tool will have difficulties. As noted in Sect. \ref{secsyn}, long-lived active regions will appear to be coherent if the time span of the data is not long enough. Another difficult case could be when the planet period is very close to the rotation period or its corresponding harmonics. These cases are amongst the trickiest of all. Good activity tracers and/or  photometric data combined with an extensive Gaussian process regression analysis \citep[e.g. ][]{Hay14,Raj15} will be required in those cases to distinguish between the periodic and quasi-periodic signals.

Analysing large amounts of data takes time. The relative speed of this tool helps to get a quick look before undertaking a computationally heavy analysis. The speed of the tool depends on many variables, such as the amount of observations, the number of scanned frequencies, the timespan of the data, and the number of trial values chosen to represent the white noise . Still, our tool is able to provide quick results. As an example, the figures in this paper were made on an Intel Core i5-480M CPU. It took respectively 23 and 26 seconds to create Figs. \ref{FigSynTest} and \ref{FigSynSNR}.

The visualisation and speed are key benefits for the tool we present here. It will be useful as an additional test to determine the nature of periodic signals in radial velocity datasets. This way we will be able to find and characterise the smallest planets.

\begin{acknowledgements}
We thank the referee for a timely and constructive report that helped improve this paper.
We would like to thank David Phillips and the members of the HARPS-N solar telescope team for sharing the data with us before their publication. The research leading to these results received funding from the European Union Seventh Framework Programme (FP7/2007-2013) under grant agreement number 313014 (ETAEARTH). ACC acknowledges support from STFC grant ST/M001296/1.
\end{acknowledgements}


\bibliographystyle{aa} 
\bibliography{References.bib} 

\begin{thebibliography}{42}
\expandafter\ifx\csname natexlab\endcsname\relax\def\natexlab#1{#1}\fi

\bibitem[{{Aigrain} {et~al.}(2012){Aigrain}, {Pont}, \& {Zucker}}]{Aig12}
{Aigrain}, S., {Pont}, F., \& {Zucker}, S. 2012, \mnras, 419, 3147

\bibitem[{{Anglada-Escud{\'e}} \& {Tuomi}(2015)}]{Angl15}
{Anglada-Escud{\'e}}, G. \& {Tuomi}, M. 2015, Science, 347, 1080

\bibitem[{{Baluev}(2013)}]{Balu13b}
{Baluev}, R.~V. 2013, \mnras, 429, 2052

\bibitem[{{Barros} {et~al.}(2014){Barros}, {Almenara}, {Deleuil}, {Diaz},
  {Csizmadia}, {Cabrera}, {Chaintreuil}, {Collier Cameron}, {Hatzes},
  {Haywood}, {Lanza}, {Aigrain}, {Alonso}, {Bord{\'e}}, {Bouchy}, {Deeg},
  {Erikson}, {Fridlund}, {Grziwa}, {Gandolfi}, {Guillot}, {Guenther}, {Leger},
  {Moutou}, {Ollivier}, {Pasternacki}, {P{\"a}tzold}, {Rauer}, {Rouan},
  {Santerne}, {Schneider}, \& {Wuchterl}}]{Barr14}
{Barros}, S.~C.~C., {Almenara}, J.~M., {Deleuil}, M., {et~al.} 2014, \aap, 569,
  A74

\bibitem[{{Boisse} {et~al.}(2012){Boisse}, {Bonfils}, \& {Santos}}]{Boi12}
{Boisse}, I., {Bonfils}, X., \& {Santos}, N.~C. 2012, \aap, 545, A109

\bibitem[{{Bonfils} {et~al.}(2005){Bonfils}, {Forveille}, {Delfosse}, {Udry},
  {Mayor}, {Perrier}, {Bouchy}, {Pepe}, {Queloz}, \& {Bertaux}}]{Bon05b}
{Bonfils}, X., {Forveille}, T., {Delfosse}, X., {et~al.} 2005, \aap, 443, L15

\bibitem[{{Bruch}(2014)}]{Bru14}
{Bruch}, A. 2014, \aap, 566, A101

\bibitem[{{Buchhave} {et~al.}(2016){Buchhave}, {Dressing}, {Dumusque}, {Rice},
  {Vanderburg}, {Mortier}, {Lopez-Morales}, {Lopez}, {Lundkvist}, {Kjeldsen},
  {Affer}, {Bonomo}, {Charbonneau}, {Collier Cameron}, {Cosentino}, {Figueira},
  {Fiorenzano}, {Harutyunyan}, {Haywood}, {Johnson}, {Latham}, {Lovis},
  {Malavolta}, {Mayor}, {Micela}, {Molinari}, {Motalebi}, {Nascimbeni}, {Pepe},
  {Phillips}, {Piotto}, {Pollacco}, {Queloz}, {Sasselov}, {S{\'e}gransan},
  {Sozzetti}, {Udry}, \& {Watson}}]{Buch16}
{Buchhave}, L.~A., {Dressing}, C.~D., {Dumusque}, X., {et~al.} 2016, \aj, 152,
  160

\bibitem[{{Cosentino} {et~al.}(2012){Cosentino}, {Lovis}, {Pepe}, {Collier
  Cameron}, {Latham}, {Molinari}, {Udry}, {Bezawada}, {Black}, {Born},
  {Buchschacher}, {Charbonneau}, {Figueira}, {Fleury}, {Galli}, {Gallie},
  {Gao}, {Ghedina}, {Gonzalez}, {Gonzalez}, {Guerra}, {Henry}, {Horne},
  {Hughes}, {Kelly}, {Lodi}, {Lunney}, {Maire}, {Mayor}, {Micela}, {Ordway},
  {Peacock}, {Phillips}, {Piotto}, {Pollacco}, {Queloz}, {Rice}, {Riverol},
  {Riverol}, {San Juan}, {Sasselov}, {Segransan}, {Sozzetti}, {Sosnowska},
  {Stobie}, {Szentgyorgyi}, {Vick}, \& {Weber}}]{Cos12}
{Cosentino}, R., {Lovis}, C., {Pepe}, F., {et~al.} 2012, in \procspie, Vol.
  8446, Ground-based and Airborne Instrumentation for Astronomy IV, 84461V

\bibitem[{{Dawson} \& {Fabrycky}(2010)}]{Daw10}
{Dawson}, R.~I. \& {Fabrycky}, D.~C. 2010, \apj, 722, 937

\bibitem[{{Dumusque} {et~al.}(2014){Dumusque}, {Boisse}, \& {Santos}}]{Dum14b}
{Dumusque}, X., {Boisse}, I., \& {Santos}, N.~C. 2014, \apj, 796, 132

\bibitem[{{Dumusque} {et~al.}(2016){Dumusque}, {Borsa}, {Damasso}, {Diaz},
  {Gregory}, {Hara}, {Hatzes}, {Rajpaul}, {Tuomi}, {Aigrain}, {Anglada-Escude},
  {Bonomo}, {Boue}, {Dauvergne}, {Frustagli}, {Giacobbe}, {Haywood}, {Jones},
  {Pinamonti}, {Poretti}, {Rainer}, {Segransan}, {Sozzetti}, \& {Udry}}]{Dum17}
{Dumusque}, X., {Borsa}, F., {Damasso}, M., {et~al.} 2016, ArXiv e-prints
  [\eprint[arXiv]{1609.03674}]

\bibitem[{{Dumusque} {et~al.}(2015){Dumusque}, {Glenday}, {Phillips},
  {Buchschacher}, {Collier Cameron}, {Cecconi}, {Charbonneau}, {Cosentino},
  {Ghedina}, {Latham}, {Li}, {Lodi}, {Lovis}, {Molinari}, {Pepe}, {Udry},
  {Sasselov}, {Szentgyorgyi}, \& {Walsworth}}]{Dum15}
{Dumusque}, X., {Glenday}, A., {Phillips}, D.~F., {et~al.} 2015, \apjl, 814,
  L21

\bibitem[{{Dumusque} {et~al.}(2012){Dumusque}, {Pepe}, {Lovis},
  {S{\'e}gransan}, {Sahlmann}, {Benz}, {Bouchy}, {Mayor}, {Queloz}, {Santos},
  \& {Udry}}]{Dum12}
{Dumusque}, X., {Pepe}, F., {Lovis}, C., {et~al.} 2012, \nat, 491, 207

\bibitem[{{Forveille} {et~al.}(2011){Forveille}, {Bonfils}, {Delfosse},
  {Alonso}, {Udry}, {Bouchy}, {Gillon}, {Lovis}, {Neves}, {Mayor}, {Pepe},
  {Queloz}, {Santos}, {Segransan}, {Almenara}, {Deeg}, \& {Rabus}}]{Forv11}
{Forveille}, T., {Bonfils}, X., {Delfosse}, X., {et~al.} 2011, ArXiv e-prints
  [\eprint[arXiv]{1109.2505}]

\bibitem[{{Gregory}(2016)}]{Greg16}
{Gregory}, P.~C. 2016, \mnras, 458, 2604

\bibitem[{{Hatzes}(2013)}]{Hat13}
{Hatzes}, A.~P. 2013, Astronomische Nachrichten, 334, 616

\bibitem[{{Hatzes} {et~al.}(2010){Hatzes}, {Dvorak}, {Wuchterl}, {Guterman},
  {Hartmann}, {Fridlund}, {Gandolfi}, {Guenther}, \& {P{\"a}tzold}}]{Hat10}
{Hatzes}, A.~P., {Dvorak}, R., {Wuchterl}, G., {et~al.} 2010, \aap, 520, A93

\bibitem[{{Haywood} {et~al.}(2014){Haywood}, {Collier Cameron}, {Queloz},
  {Barros}, {Deleuil}, {Fares}, {Gillon}, {Lanza}, {Lovis}, {Moutou}, {Pepe},
  {Pollacco}, {Santerne}, {S{\'e}gransan}, \& {Unruh}}]{Hay14}
{Haywood}, R.~D., {Collier Cameron}, A., {Queloz}, D., {et~al.} 2014, \mnras,
  443, 2517

\bibitem[{{Howard} {et~al.}(2011){Howard}, {Johnson}, {Marcy}, {Fischer},
  {Wright}, {Henry}, {Isaacson}, {Valenti}, {Anderson}, \& {Piskunov}}]{How11}
{Howard}, A.~W., {Johnson}, J.~A., {Marcy}, G.~W., {et~al.} 2011, \apj, 726, 73

\bibitem[{{Hu{\'e}lamo} {et~al.}(2008){Hu{\'e}lamo}, {Figueira}, {Bonfils},
  {Santos}, {Pepe}, {Gillon}, {Azevedo}, {Barman}, {Fern{\'a}ndez}, {di Folco},
  {Guenther}, {Lovis}, {Melo}, {Queloz}, \& {Udry}}]{Hue08}
{Hu{\'e}lamo}, N., {Figueira}, P., {Bonfils}, X., {et~al.} 2008, \aap, 489, L9

\bibitem[{{Lomb}(1976)}]{Lomb76}
{Lomb}, N.~R. 1976, \apss, 39, 447

\bibitem[{{Mayor} {et~al.}(2009){Mayor}, {Bonfils}, {Forveille}, {Delfosse},
  {Udry}, {Bertaux}, {Beust}, {Bouchy}, {Lovis}, {Pepe}, {Perrier}, {Queloz},
  \& {Santos}}]{Mayor09}
{Mayor}, M., {Bonfils}, X., {Forveille}, T., {et~al.} 2009, \aap, 507, 487

\bibitem[{{Mayor} {et~al.}(2003){Mayor}, {Pepe}, {Queloz}, {Bouchy},
  {Rupprecht}, {Lo Curto}, {Avila}, {Benz}, {Bertaux}, {Bonfils}, {Dall},
  {Dekker}, {Delabre}, {Eckert}, {Fleury}, {Gilliotte}, {Gojak}, {Guzman},
  {Kohler}, {Lizon}, {Longinotti}, {Lovis}, {Megevand}, {Pasquini}, {Reyes},
  {Sivan}, {Sosnowska}, {Soto}, {Udry}, {van Kesteren}, {Weber}, \&
  {Weilenmann}}]{Mayor03}
{Mayor}, M., {Pepe}, F., {Queloz}, D., {et~al.} 2003, The Messenger, 114, 20

\bibitem[{{Mortier} {et~al.}(2015){Mortier}, {Faria}, {Correia}, {Santerne}, \&
  {Santos}}]{ME15}
{Mortier}, A., {Faria}, J.~P., {Correia}, C.~M., {Santerne}, A., \& {Santos},
  N.~C. 2015, \aap, 573, A101

\bibitem[{{Phillips} {et~al.}(2016){Phillips}, {Glenday}, {Dumusque},
  {Buchschacher}, {Cameron}, {Cecconi}, {Charbonneau}, {Cosentino}, {Ghedina},
  {Haywood}, {Latham}, {Li}, {Lodi}, {Lovis}, {Molinari}, {Pepe}, {Sasselov},
  {Szentgyorgyi}, {Udry}, \& {Walsworth}}]{Phi16}
{Phillips}, D.~F., {Glenday}, A.~G., {Dumusque}, X., {et~al.} 2016, in
  \procspie, Vol. 9912, Society of Photo-Optical Instrumentation Engineers
  (SPIE) Conference Series, 99126Z

\bibitem[{{Phillips et al.}(2017)}]{Phi17}
{Phillips et al.}, D. 2017, in prep., in prep.

\bibitem[{{Queloz} {et~al.}(2009){Queloz}, {Bouchy}, {Moutou}, {Hatzes},
  {H{\'e}brard}, {Alonso}, {Auvergne}, {Baglin}, {Barbieri}, {Barge}, {Benz},
  {Bord{\'e}}, {Deeg}, {Deleuil}, {Dvorak}, {Erikson}, {Ferraz Mello},
  {Fridlund}, {Gandolfi}, {Gillon}, {Guenther}, {Guillot}, {Jorda}, {Hartmann},
  {Lammer}, {L{\'e}ger}, {Llebaria}, {Lovis}, {Magain}, {Mayor}, {Mazeh},
  {Ollivier}, {P{\"a}tzold}, {Pepe}, {Rauer}, {Rouan}, {Schneider},
  {Segransan}, {Udry}, \& {Wuchterl}}]{Que09}
{Queloz}, D., {Bouchy}, F., {Moutou}, C., {et~al.} 2009, \aap, 506, 303

\bibitem[{{Queloz} {et~al.}(2001){Queloz}, {Henry}, {Sivan}, {Baliunas},
  {Beuzit}, {Donahue}, {Mayor}, {Naef}, {Perrier}, \& {Udry}}]{Que01}
{Queloz}, D., {Henry}, G.~W., {Sivan}, J.~P., {et~al.} 2001, \aap, 379, 279

\bibitem[{{Rajpaul} {et~al.}(2015){Rajpaul}, {Aigrain}, {Osborne}, {Reece}, \&
  {Roberts}}]{Raj15}
{Rajpaul}, V., {Aigrain}, S., {Osborne}, M.~A., {Reece}, S., \& {Roberts}, S.
  2015, \mnras, 452, 2269

\bibitem[{{Rajpaul} {et~al.}(2016){Rajpaul}, {Aigrain}, \& {Roberts}}]{Raj16}
{Rajpaul}, V., {Aigrain}, S., \& {Roberts}, S. 2016, \mnras, 456, L6

\bibitem[{{Ramsay} {et~al.}(2016){Ramsay}, {Hakala}, {Wood}, {Howell}, {Smale},
  {Still}, \& {Barclay}}]{Ram16}
{Ramsay}, G., {Hakala}, P., {Wood}, M.~A., {et~al.} 2016, \mnras, 455, 2772

\bibitem[{{Roberts} {et~al.}(1987){Roberts}, {Lehar}, \& {Dreher}}]{Rob87}
{Roberts}, D.~H., {Lehar}, J., \& {Dreher}, J.~W. 1987, \aj, 93, 968

\bibitem[{{Robertson} {et~al.}(2014){Robertson}, {Mahadevan}, {Endl}, \&
  {Roy}}]{Rob14b}
{Robertson}, P., {Mahadevan}, S., {Endl}, M., \& {Roy}, A. 2014, Science, 345,
  440

\bibitem[{{Robertson} {et~al.}(2015){Robertson}, {Roy}, \&
  {Mahadevan}}]{Rob15b}
{Robertson}, P., {Roy}, A., \& {Mahadevan}, S. 2015, \apjl, 805, L22

\bibitem[{{Santos} {et~al.}(2014){Santos}, {Mortier}, {Faria}, {Dumusque},
  {Adibekyan}, {Delgado-Mena}, {Figueira}, {Benamati}, {Boisse}, {Cunha},
  {Gomes da Silva}, {Lo Curto}, {Lovis}, {Martins}, {Mayor}, {Melo}, {Oshagh},
  {Pepe}, {Queloz}, {Santerne}, {S{\'e}gransan}, {Sozzetti}, {Sousa}, \&
  {Udry}}]{San14}
{Santos}, N.~C., {Mortier}, A., {Faria}, J.~P., {et~al.} 2014, \aap, 566, A35

\bibitem[{{Scargle}(1982)}]{Sca82}
{Scargle}, J.~D. 1982, \apj, 263, 835

\bibitem[{{Su{\'a}rez Mascare{\~n}o} {et~al.}(2016){Su{\'a}rez Mascare{\~n}o},
  {Gonz{\'a}lez Hern{\'a}ndez}, {Rebolo}, {Astudillo-Defru}, {Bonfils},
  {Bouchy}, {Delfosse}, {Forveille}, {Lovis}, {Mayor}, {Murgas}, {Pepe},
  {Santos}, {Udry}, {W{\"u}nsche}, \& {Velasco}}]{Sua16}
{Su{\'a}rez Mascare{\~n}o}, A., {Gonz{\'a}lez Hern{\'a}ndez}, J.~I., {Rebolo},
  R., {et~al.} 2016, ArXiv e-prints [\eprint[arXiv]{1611.02122}]

\bibitem[{{Udry} {et~al.}(2007){Udry}, {Bonfils}, {Delfosse}, {Forveille},
  {Mayor}, {Perrier}, {Bouchy}, {Lovis}, {Pepe}, {Queloz}, \&
  {Bertaux}}]{Udry07b}
{Udry}, S., {Bonfils}, X., {Delfosse}, X., {et~al.} 2007, \aap, 469, L43

\bibitem[{{Vogt} {et~al.}(1994){Vogt}, {Allen}, {Bigelow}, {Bresee}, {Brown},
  {Cantrall}, {Conrad}, {Couture}, {Delaney}, {Epps}, {Hilyard}, {Hilyard},
  {Horn}, {Jern}, {Kanto}, {Keane}, {Kibrick}, {Lewis}, {Osborne},
  {Pardeilhan}, {Pfister}, {Ricketts}, {Robinson}, {Stover}, {Tucker}, {Ward},
  \& {Wei}}]{Vogt94}
{Vogt}, S.~S., {Allen}, S.~L., {Bigelow}, B.~C., {et~al.} 1994, in Society of
  Photo-Optical Instrumentation Engineers (SPIE) Conference Series, Vol. 2198,
  Society of Photo-Optical Instrumentation Engineers (SPIE) Conference Series,
  ed. {D.~L.~Crawford \& E.~R.~Craine}, 362--+

\bibitem[{{Vogt} {et~al.}(2010){Vogt}, {Butler}, {Rivera}, {Haghighipour},
  {Henry}, \& {Williamson}}]{Vogt10}
{Vogt}, S.~S., {Butler}, R.~P., {Rivera}, E.~J., {et~al.} 2010, \apj, 723, 954

\bibitem[{{Zechmeister} \& {K{\"u}rster}(2009)}]{Zech09}
{Zechmeister}, M. \& {K{\"u}rster}, M. 2009, \aap, 496, 577

\end{thebibliography}

\end{document}